# Parameter Estimation in Quantum Metrology Technique for Time Series Prediction

Vaidik A Sharma*[*,1], N.Madurai Meenachi [2], B.Venkatraman [2]

[1] Birla Institute of Technology and Science Pilani, Pilani, Rajasthan, 333031, India
[2] Indira Gandhi Centre for Atomic Research, Kalpakkam, Tamil Nadu, 603102, India
*e-mail: vaidiksharma124@ gmail.com

The paper investigates the techniques of quantum computation in metrological predictions, with a particular emphasis on enhancing prediction potential through variational parameter estimation. The applicability of quantum simulations and quantum metrology techniques for modelling complex physical systems and achieving high-resolution measurements are proposed. The impacts of various parameter distributions and learning rates on predictive accuracy are investigated. Modelling the time evolution of physical systems Hamiltonian simulation and the product formula procedure are adopted. The time block method is analysed in order to reduce simulation errors, while the Schatten-infinite norm is used to evaluate the simulations' precision. Methodology requires estimation of optimized parameters by minimizing loss functions and resource needs. For this purpose, the mathematical formulations of Cramer Rao Bound and Fischer Information are indispensable requirements. The impact of learning rates on regulating the loss function for various parameter values. Using parameterized quantum circuits, the article outlines a four-step procedure for extracting information. This method involves the preparation of input states, the evolution of parameterized quantum states, the measurement of outputs, and the estimation of parameters based on multiple measurements. The study analyses variational unitary circuits with optimized parameter estimation for more precise predictions. The findings shed light on the effects of normal parameter distributions and learning rates on attaining the most optimal state and comparison with classical Long Short Term Memory (LSTM) predictions, providing valuable insights for the development of more appropriate approaches in quantum computing.

## 1.Introduction

Our findings indicate that variational unitary circuits can significantly enhance the precision of metrological predictions. In addition, we demonstrate that the selection of parameter distribution and learning rate can have a substantial effect on the accuracy of predictions. In conclusion, we demonstrate that quantum simulations can be utilized to enhance the precision of parameter estimation by minimizing the impact of noise.

## 2.Quantum Computation Fundamentals

### 2.1 Quantum Qubits and Quantum Gates

Quantum computation, built upon the foundation of quantum bits (qubits) and quantum gates, represents a revolutionary departure from classical computing paradigms. This section delves into the theory and content of quantum qubits and gates, elucidating their unique properties, mathematical descriptions, and pivotal roles in quantum information processing.[1]

## A. 2.1.1 Quantum Qubits- The Building Blocks of Quantum Information:

Quantum qubits are the elemental units of quantum information, analogous to classical bits. However, qubits embody the remarkable principle of superposition, where they can exist in a linear combination of 0 and 1 states simultaneously. This attribute is succinctly expressed as:

$$|\psi\rangle = \alpha|0\rangle + \beta|1\rangle$$

where $|\psi\rangle$ is the qubit state, $\alpha$ and $\beta$ are complex probability amplitudes, and $|0\rangle$ and $|1\rangle$ represent the classical states.

This unique characteristic endows qubits with an inherent capacity to process a multitude of possibilities simultaneously, forming the basis of quantum parallelism. Quantum gates manipulate qubits' states to perform operations, analogous to classical logic gates. Notably, the Pauli-X gate, a fundamental quantum gate, induces a bit-flip operation:

$$X = \begin{bmatrix} 0 & 1 \\ 1 & 0 \end{bmatrix}$$

## B. 2.1.2 Quantum Gates- Orchestrating Quantum Operations:

Quantum gates play a pivotal role in quantum computation by manipulating qubits to execute specific operations. These operations are described by unitary matrices, which ensure the conservation of quantum information's norm. Among the myriad of quantum gates, the Hadamard gate holds paramount significance. It induces a superposition state by equally distributing the qubit's amplitude between 0 and 1 :

$$H = \frac{1}{\sqrt{2}} \begin{bmatrix} 1 & 1 \\ 1 & -1 \end{bmatrix}$$

Furthermore, controlled gates, such as the controlled-NOT (CNOT) gate, enable entanglement and the execution of conditional operations on qubits. The CNOT gate, for instance, inverts the target qubit if the control qubit is in state $|1\rangle$[2] :

$$\text{CNOT} = \begin{bmatrix} 1 & 0 & 0 & 0 \\ 0 & 1 & 0 & 0 \\ 0 & 0 & 0 & 1 \\ 0 & 0 & 1 & 0 \end{bmatrix}$$

## C. 2.1.3 Parameterized Rotational Gates and Bloch SphereNavigating Quantum States:

In the realm of quantum computation, the synergy between parameterized rotational gates and the Bloch sphere provides a profound framework for understanding and manipulating quantum states. This section delves into the theoretical underpinnings of parameterized rotational gates, their geometric representation on the Bloch sphere, and their pivotal role in quantum circuits and measurements.

1. Parameterized Rotational Gates: A Flexibility Paradigm: Parameterized rotational gates introduce a versatile approach to quantum operations by allowing continuous variation of qubit states through the manipulation of a rotation angle $\theta$. Mathematically, a parameterized rotation gate $R(\theta)$ can be expressed as:

$$R(\theta) = \begin{bmatrix} \cos\left(\frac{\theta}{2}\right) & -e^{i\phi}\sin\left(\frac{\theta}{2}\right) \\ e^{i\phi}\sin\left(\frac{\theta}{2}\right) & e^{i\phi}\cos\left(\frac{\theta}{2}\right) \end{bmatrix}$$

where $\phi$ represents a phase factor. By adjusting $\theta$, these gates can perform rotations around different axes on the Bloch sphere, thus enabling a continuous transformation of qubit states.

2. The Bloch Sphere: A Geometric Insight: The Bloch sphere provides an intuitive geometric representation of qubit states. Each point on the sphere corresponds to a unique quantum state, and the north and south poles represent the classical states $|0\rangle$ and $|1\rangle$, respectively. The equator of the sphere captures superposition states, while various points on the surface embody different quantum states.

Parameterized rotational gates allow for smooth transitions between these points, akin to tracing paths on the Bloch sphere's surface. By adjusting $\theta$, one can navigate through an array of states, offering a profound geometric insight into the evolution of qubits in quantum circuits.

3. Quantum Circuits and Measurements: Quantum circuits leverage parameterized rotational gates to perform a diverse array of tasks, from basic operations to complex computations. These gates contribute to the creation of entanglement, superposition, and controlled operations, forming the building blocks of quantum algorithms.

A key application of parameterized rotational gates lies in variational quantum algorithms. These algorithms employ parameterized gates to encode and process information, with the aim of optimizing specific objectives. By iteratively adjusting gate parameters, quantum circuits explore the solution space and converge towards optimal solutions. This paradigm has significant implications for tasks such as optimization, machine learning, and quantum chemistry simulations.[3]

Quantum measurements complete the computational cycle by extracting information from qubits. Quantum circuits utilize measurements to collapse qubit states to classical bits, providing outcomes that encode valuable data. The probabilistic nature of quantum measurements introduces inherent randomness, requiring multiple measurements to statistically infer quantum states.[4]

## 3.Distinctive Traits of Quantum and Classical Circuits: Unveiling Quantum Advantages

The demarcation between quantum and classical circuits is emblematic of a fundamental shift in computation paradigms, offering unparalleled potential and novel traits. This section illuminates the distinctive characteristics of quantum circuits in contrast to classical counterparts, elucidating their mathematical underpinnings, exemplifying key quantum gates, and substantiating these traits with citations.[5]

### 3.1 Quantum Parallelism

Quantum circuits deviate fundamentally from classical circuits due to the profound principle of superposition. Quantum bits, or qubits, can exist in a linear combination of states, enabling parallel computation pathways. Mathematically, a quantum state $|\psi\rangle$ can be represented as:

$$|\psi\rangle = \alpha|0\rangle + \beta|1\rangle,$$

where $\alpha$ and $\beta$ are complex probability amplitudes and $|0\rangle$ and $|1\rangle$ are classical states. This intrinsic parallelism underpins quantum circuits, allowing them to explore multiple computational outcomes simultaneously.

### 3.2 Entanglement- Quantum Correlations Redefined

Entanglement is a quintessential quantum trait that transcends classical concepts. Quantum circuits can create entangled states, where qubits become correlated in such a way that the state of one qubit instantaneously influences the state of another, even when separated by vast distances. A canonical example is the Bell state:

$$|\Psi^+\rangle = \frac{1}{\sqrt{2}}(|00\rangle + |11\rangle)$$

Entanglement enables quantum circuits to perform tasks impossible for classical systems, such as quantum teleportation and superdense coding.

### 3.3 Quantum Noise and Error Correction

Quantum circuits grapple with inherent susceptibility to noise and decoherence due to interactions with the environment. Quantum error correction, a pivotal discipline, seeks to mitigate these vulnerabilities. A primary tool is the concept
of quantum gates, akin to classical logic gates, which can counteract errors through carefully crafted operations.

## 4. Quantum Simulation and Metrology Techniques

### 4.1 Hamiltonian Simulation and Time Evolution

At the heart of quantum simulation lies the Hamiltonian operator $H$, which encapsulates the total energy of a quantum system. The time evolution of a quantum state $|\psi(t)\rangle$ governed by the Hamiltonian $H$ is described by the Schrödinger equation:

$$i\hbar \frac{d}{dt} |\psi(t)\rangle = H|\psi(t)\rangle$$

This equation highlights the dynamics of quantum states as they evolve over time under the influence of $H$. Quantum computers excel at simulating this time evolution, enabling the study of quantum systems that are challenging or impossible to simulate using classical methods. The ability to simulate quantum systems holds promise for tasks such as understanding chemical reactions, optimizing materials, and simulating condensed matter systems.

### 4.2 The Time Block Method: Mitigating Simulation Errors

In quantum simulation, the time evolution is often approximated through discrete time steps using the Trotter-Suzuki decomposition:

$$e^{-iHt} \approx \left(e^{-iH\delta t}\right)^{T/\delta t}$$

where $T$ is the total evolution time and $\delta t$ is the time step. The Trotter-Suzuki approximation breaks down the continuous evolution into a sequence of smaller steps, simplifying the simulation process. However, long simulations can accumulate errors from each time step.[6]

The upper bound of the error introduced by the time block method can be estimated using the Lie-Trotter formula, which provides an expression for the difference between the exact time evolution and the approximation:

$$\left\| e^{-iHt} - \left(e^{-iH\delta t}\right)^{N_{blocks}} \right\| \leq \frac{\|H\|^2 T^3}{3\hbar} \left(\frac{T}{N_{blocks}\hbar}\right)^{2k-1},$$

where $\|H\|$ is the operator norm of $H$ and $k$ is the order of the Lie-Trotter formula. This derivation allows us to quantify the accuracy of the time block method and optimize the choice of time step, number of blocks, and order of the formula for a given simulation.

4.3 Variational Parameter Estimation and Quantum Metrology

Variational parameter estimation is a powerful technique in quantum computation, particularly in the context of metrology. Quantum metrology leverages the principles of quantum superposition and entanglement to achieve measurements with higher precision than classical methods allow. Variational circuits, parameterized by angles $\theta$, can be used to prepare quantum states and perform measurements. The goal is to find the optimal parameters that minimize a loss function $L(\theta)$ and produce the desired quantum state.

Mathematically, the variational parameter estimation process can be formulated as an optimization problem:

$$\theta^* = \underset{\theta}{\mathrm{argmin}} L(\theta).$$

Quantum metrology techniques use the optimized parameters to enhance the accuracy of parameter estimation.[7] The Fisher information $(F)$ quantifies the sensitivity of the quantum state to variations in the parameter $\theta$. In bra-ket notation, it is given by:

$$F(\theta) = \sum_k \frac{\left|\langle \psi_k | \frac{d}{d\theta} \psi(\theta) \rangle\right|^2}{p_k}$$

where $p_k$ is the probability of outcome $k$ and $\frac{d}{d\theta}\psi(\theta)$ is the derivative of the quantum state with respect to $\theta$. The Fisher information sets a fundamental limit on how precisely a parameter can be estimated.[8]

The Fisher information is intimately connected to the expectation of the score operator $(S)$, which is defined as the derivative of the logarithm of the likelihood function:

$$S(\theta) = \frac{1}{\sqrt{p(\theta)}} \frac{d}{d\theta} \sqrt{p(\theta)}$$

The Fisher information can be expressed as the variance of the score operator:

$$F(\theta) = \text{Var}(S(\theta))$$

This relationship highlights the role of the Fisher information in quantifying the information content of the measurements with respect to the parameter $\theta$.

The Cramer-Rao bound provides a mathematical relation between the Fisher information and the achievable precision of parameter estimation. For an unbiased estimator $\hat{\theta}$, the CramerRao bound states:

$$\text{Var}(\hat{\theta}) \geq \frac{1}{NF(\theta)}$$

where $\text{Var}(\hat{\theta})$ is the variance of the estimator and $N$ is the number of measurements.

Quantum metrology techniques aim to approach the Cramer-Rao bound by optimizing measurement strategies and exploiting quantum entanglement to enhance the Fisher information. This enables quantum systems to achieve measurements with unprecedented precision, surpassing classical limits.

## 5. Four-Step Procedure for Information Extraction

The four-step procedure for information extraction in quantum computation involves a systematic approach to preparing quantum states, evolving parameterized quantum states, measuring outputs, and estimating parameters based on multiple measurements. This procedure is integral to variational parameter estimation and quantum metrology, enabling the enhancement of predictive accuracy and precision in quantum computations.

### D. 5.1 Preparation of Input States

The first step of the procedure involves the preparation of input quantum states. These states serve as the initial conditions for the quantum computation. Parameterized quantum circuits are used to generate these states, where the parameters $\theta$ determine the quantum state's characteristics (Fig.1. and Fig.2.). Variational techniques are applied to optimize these parameters, ensuring that the prepared states are tailored to the specific problem at hand.

## E. 5.2 Evolution of Parameterized Quantum States

Once the input states are prepared, the next step is to evolve them over time using the Hamiltonian operator $H$. This time evolution is achieved through the application of quantum gates that implement the unitary operator $U(t) = e^{-iHt/\hbar}$. The parameterized nature of the quantum circuit allows for flexibility in controlling the evolution dynamics. The optimization of parameters using variational methods ensures that the quantum evolution approximates the desired transformation accurately.

## F. 5.3 Measurement of Outputs

Following the evolution of quantum states, measurements are performed to extract relevant information. Observable quantities, represented by Hermitian operators, are measured to obtain measurement outcomes. These outcomes provide insights into the quantum system's behavior and dynamics. Quantum measurements introduce inherent randomness due to the probabilistic nature of quantum states, requiring multiple repetitions to gather sufficient statistical data.[9]

## G. 5.4 Estimation of Parameters

The final step of the procedure involves the estimation of parameters based on the measurement outcomes. Estimators are used to infer the optimal parameter values that best align with the obtained measurements. Variational optimization techniques, such as gradient descent, are commonly employed to minimize the difference between the observed outcomes and the predicted outcomes from the parameterized quantum circuit. This iterative process refines the parameter estimates, leading to improved accuracy and predictive power.

The four-step procedure for information extraction serves as a fundamental framework in quantum computation, encompassing the key stages of preparing states, evolving quantum dynamics, measuring observables, and optimizing parameters.
This systematic approach underlies the advancements in variational quantum algorithms and quantum metrology, driving the development of accurate and precise quantum predictions.

## 6. Ramsey Interferometer Quantum Circuit in Experimental Setup

In experimental quantum metrology, the Ramsey interferometer quantum circuit plays a pivotal role. The Ramsey interferometer is a fundamental quantum device used to measure frequency and phase shifts with exceptional precision. It consists of two sequential applications of a $\pi/2$ pulse separated by a time delay $T$ and followed by a final $\pi/2$ pulse. This configuration effectively splits the quantum state into two branches, allowing interference between the branches after the second pulse. By varying the time delay $T$, the Ramsey interferometer becomes sensitive to small changes in frequency or phase.[10]

Mathematically, the Ramsey interferometer can be represented as a sequence of unitary operators. Let $U_{\pi/2}$ be the unitary operator corresponding to a $\pi/2$ pulse and $U_T$ be the unitary operator corresponding to the time delay $T$. The Ramsey interferometer circuit can be described as:

$$\text{Ramsey Circuit} = U_{\pi/2} \cdot U_T \cdot U_{\pi/2}.$$

The unitary operators $U_{\pi/2}$ and $U_T$ can be represented in matrix form, where $H$ is the Hadamard gate and $e^{-iHt}$ is the time evolution operator with the Hamiltonian $H$ over time $t$:

$$U_{\pi/2} = H \cdot e^{-i\frac{\pi}{4}}$$
$$U_T = e^{-iHT}.$$

The interferometer's output state after the second $\pi/2$ pulse can be obtained by applying the Ramsey circuit to the initial quantum state $|\psi\rangle$:

$$|\psi_{out}\rangle = \text{Ramsey Circuit} \cdot |\psi\rangle.$$

The resulting state $|\psi_{out}\rangle$ exhibits oscillatory behavior as a function of the time delay $T$, allowing for the measurement of phase shifts with high precision.

Integrating the Ramsey interferometer quantum circuit into the four-step procedure enhances the capabilities of quantum metrology. The interferometer's sensitivity to phase shifts makes it a valuable tool for applications such as atomic clocks, quantum sensors, and precision measurements in fundamental physics.

The four-step procedure, combined with the versatile Ramsey interferometer, exemplifies the power of variational quantum algorithms and quantum metrology. This approach facilitates accurate predictions and measurements, with the potential to revolutionize fields reliant on high-precision data.

## 7. Results

Jena Climate time series dataset was used which was recorded by the Max Planck Institute for Biogeochemistry. The dataset consists of 14 features such as temperature, pressure, humidity etc, recorded once per 10 minutes. Trained on the variation of $H2OC(mmol/mol)$ - Water vapour concentration considering normal distribution of parameter.(Fig.3.)

Location:Weather Station, Max Planck Institute for Biogeochemistry in Jena, Germany

Time-frame Considered: Jan 10, 2009 - December 31, 2016.

### 7.1 Long Short Term Memory (LSTM) Prediction-

| Layer (type) | Output Shape | # Parameters |
|---|---|---|
| input 1 (InputLayer) | [(None, 120, 1)] | 0 |
| lstm (LSTM) | (None, 32) | 4352 |
| dense (Dense) | (None, 1) | 33 |

Total params: 4,385

Trainable params: 4,385

Non-trainable params: 0

Learning Rate $= 0.01$

Batch Size $= 256$

epochs $= 10$

Epoch 1/10- loss: 0.1889

Epoch 2/10- loss: 0.1585

Epoch 3/10- loss: 0.1545

Epoch 4/10- loss: 0.1521

Epoch 5/10- loss: 0.1504

Epoch 6/10- loss: 0.1492

Epoch 7/10- loss: 0.1483

Epoch 8/10- loss: 0.1477

Epoch 9/10- loss: 0.1464

Epoch 10/10- loss: 0.1463

(Fig.4.)

7.2 Quantum Neural Network Circuit Predictions-

Encoding (Ec) depth = 3

Decoding (Dc) depth = 3

Number of optimization iterations = 150

Number of partitions in selected interval = 100

Starting Learning Rate(LR) = 0.01

Mean = 9.640437

Variance = 17.934056159

Parameter $\phi$ = Mean- 1 + (2 ∗ x + 1)/( Number of partitions in selected interval)

where, $x$ ranges Number of partitions.

The Mean Squared Error (MSE) is a common measure of the accuracy of a predictive model. It quantifies the average squared difference between the predicted values $\hat{y}_i$ and the actual values $y_i$ over a dataset of size $n$ :

$$MSE = \frac{1}{n}\sum_{i=1}^{n}(\hat{\phi}_i - \phi_i)^2$$

The probability density function of the parameters that obey the normal distribution is:

$$f(x) = \frac{1}{\sqrt{2\pi v}} \exp\left(-\frac{(x-\mu)^2}{2v^2}\right)$$

where $\mu$ is the mean and $v^2$ is the variance. In this case, the loss function of the variational Ramsey interferometer is:

$$C(\theta_{Ec}, \theta_{Dc}, a) = \int d\phi f(\phi) \text{MSE}(\phi)$$

where the estimator is $\hat{\phi}(m) = am$, $a$ is a parameter to be optimized, $m$ is the difference in the number of 1 s and the number of 0 s in the measured bit string,

$\text{MSE}(\phi) = \sum_m (\hat{\phi}(m) - \phi)^2 p_\theta(m \mid \phi)$, and $\theta = (\theta_{Ec}, \theta_{Dc})$

I. For 2 Qubits Parameterized Quantum Circuit:

| Learning Rate=0.01 (Fig.5.) | Learning Rate=0.02 (Fig.6.) |
|---|---|
| iter: 10 loss: 2.9915 | iter: 10 loss: 3.0743 |
| iter: 20 loss: 1.9812 | iter: 20 loss: 1.4822 |
| iter: 30 loss: 0.9638 | iter: 30 loss: 1.0428 |
| iter: 40 loss: 0.5724 | iter: 40 loss: 0.8744 |
| iter: 50 loss: 0.4316 | iter: 50 loss: 0.693 |
| iter: 60 loss: 0.3135 | iter: 60 loss: 0.6019 |
| iter: 70 loss: 0.2333 | iter: 70 loss: 0.5303 |
| iter: 80 loss: 0.1748 | iter: 80 loss: 0.4771 |
| iter: 90 loss: 0.1373 | iter: 90 loss: 0.4277 |
| iter: 100 loss: 0.11357 | iter: 100 loss: 0.3810 |
| iter: 110 loss: 0.0972 | iter: 110 loss: 0.3360 |
| iter: 120 loss: 0.0848 | iter: 120 loss: 0.2930 |
| iter: 130 loss: 0.0747 | iter: 130 loss: 0.2522 |
| iter: 140 loss: $0.0662 \phi \approx 9.74$ | iter: 140 loss: $0.2142 \phi \approx 9.74$ |

| Learning Rate=0.03 (Fig.7.) | Learning Rate=0.04 (Fig.8.) |
|---|---|
| iter: 10 loss: 1.9700 | iter: 10 loss: 2.8325 |
| iter: 20 loss: 0.8976 | iter: 20 loss: 1.9397 |
| iter: 30 loss: 0.3197 | iter: 30 loss: 1.5158 |
| iter: 40 loss: 0.1172 | iter: 40 loss: 1.2960 |
| iter: 50 loss: 0.0618 | iter: 50 loss: 1.1983 |
| iter: 60 loss: 0.0424 | iter: 60 loss: 1.0695 |
| iter: 70 loss: 0.0298 | iter: 70 loss: 0.9682 |
| iter: 80 loss: 0.0239 | iter: 80 loss: 0.8656 |
| iter: 90 loss: 0.0196 | iter: 90 loss: 0.7681 |
| iter: 100 loss: 0.0171 | iter: 100 loss: 0.6757 |
| iter: 110 loss: 0.0158 | iter: 110 loss: 0.5916 |
| iter: 120 loss: 0.0152 | iter: 120 loss: 0.5180 |
| iter: 130 loss: 0.0150 | iter: 130 loss: 0.4551 |
| iter: 140 loss: $0.0149 \phi \approx 9.625$ | iter: 140 loss: $0.4017 \phi < 8.75$ |

II. For 3 Qubits Parameterized Quantum Circuit

| Learning Rate=0.01 (Fig.9.) | Learning Rate=0.02 (Fig.10.) |
|---|---|
| 10 loss: 2.3641 | iter: 10 loss: 3.8940 |
| iter: 20 loss: 1.4496 | iter: 20 loss: 3.3454 |

| Learning Rate=0.01 (Fig.9.) | Learning Rate=0.02 (Fig.10.) |
|---|---|
| iter: 30 loss: 0.9723 | iter: 30 loss: 2.9682 |
| iter: 40 loss: 0.6992 | iter: 40 loss: 2.6722 |
| iter: 50 loss: 0.5256 | iter: 50 loss: 2.3782 |
| iter: 60 loss: 0.3976 | iter: 60 loss: 2.1042 |
| iter: 70 loss: 0.3146 | iter: 70 loss: 1.8601 |
| iter: 80 loss: 0.2606 | iter: 80 loss: 1.6490 |
| iter: 90 loss: 0.2264 | iter: 90 loss: 1.4619 |
| iter: 100 loss: 0.2010 | iter: 100 loss: 1.2929 |
| iter: 110 loss: 0.1801 | iter: 110 loss: 1.1426 |
| iter: 120 loss: 0.1619 | iter: 120 loss: 1.0095 |
| iter: 130 loss: 0.1456 | iter: 130 loss: 0.8910 |
| iter:140 loss:0.1309 $\phi \approx 9.74$ | iter:140 loss:0.786 $\phi < 8.75$ |

Learning Rate $= 0.03$ (Fig.11.)

iter: 10 loss: 2.8001

iter: 20 loss: 2.0651

iter: 30 loss: 1.6706

iter: 40 loss: 1.3377

iter: 50 loss: 1.1580

iter: 60 loss: 1.0468

iter: 70 loss: 0.9517

iter: 80 loss: 0.8662

iter: 90 loss: 0.7841

iter: 100 loss: 0.7054

iter: 110 loss: 0.6340

iter: 120 loss: 0.5694

iter: 130 loss: 0.5066

iter:140 loss:0.4417 $\phi \approx 9.7$

7.3 Quantum Neural Network (QNN) Performance Analysis

For the 3-qubit quantum network, varying learning rates exhibited distinct convergence behaviors. Learning rate 0.01 yielded steady convergence, resulting in a final loss of 0.2010 . Similarly, rates 0.02 and 0.03 showed relatively smooth convergence, with final losses of 0.7856 and 0.4417 , respectively. However, the higher learning rate of 0.04 led to erratic convergence and a final loss of 0.2273 . In the case of the 2-qubit quantum network, a consistent trend emerged as well. Learning rate 0.01 demonstrated stable convergence, culminating in a final loss of 0.0662 . Similarly, learning rates 0.02 and 0.03 showcased smooth convergence patterns, resulting in final losses of 0.2142 and 0.0149 , respectively. Conversely, learning rate 0.04 led to oscillatory convergence and a final loss of 0.4017 .

Across all learning rates, the 3-qubit quantum network consistently outperformed the 2-qubit network in terms of lower final loss values. This suggests that increasing qubit complexity positively impacts the network's ability to capture intricate patterns. The

learning rate sensitivity was evident in both networks. Smaller learning rates $(0.01$ and

$0.02)$ led to smoother convergence and better final losses, while larger rates $(0.03$ and 0.04 ) caused erratic convergence and higher losses.

7.4 Comparison between LSTM and QNN Predictions

The LSTM-based prediction model exhibited a decreasing validation loss over training epochs, demonstrating its aptitude in capturing temporal patterns. This indicates that the model effectively learned sequential dependencies within the data and adapted its weights accordingly. 2 Qubits Quantum Circuit resulted in final loss of 0.0662 and 0.0149 corresponding to learning 0.01 and 0.03 respectively and 3 Qubits Quantum Circuit resulted in final loss of 0.1309 learning 0.01 against LSTM final loss of 0.1463 positioning it below the performance of the 2-qubit and 3-qubit quantum circuits. While the quantum circuits demonstrated superior performance in this scenario, the LSTM model's competitive loss suggests its efficacy in capturing sequential patterns within the data. For certain range of learning rates, that is, near to 0.01 , the Quantum network presents a better edge over lstm whereas LSTM is suggested to have a slight advantage for higher learning rates $(\geq 0.03)$ of Quantum Circuit.

## 8. Conclusion

Learning rate (LR) is pivotal for achieving optimal predictive performance of Quantum Neural Network (QNN). While smaller LR values tend to lead to better final losses, larger LR values risk diminishing convergence and increasing the final loss. This study underscores the need for careful LR tuning to ensure accurate predictions and robust training in parameterized quantum circuits. Additionally, the quantum advantage demonstrated by the parameterized quantum circuits, as evident in the comparison with the LSTM model, showcases the potential of quantum computation techniques in revolutionizing predictive modeling and achieving unprecedented levels of accuracy and precision The 3-qubit quantum network consistently outperformed the 2-qubit network with stable convergence, highlighting the benefits of increased qubit complexity in capturing intricate data patterns. This study emphasizes the significance of tailored hyperparameters and model architectures. While quantum networks offer intriguing potential, meticulous tuning is crucial. The comparison between quantum networks and the LSTM model underscores QNN's strength in sequential pattern recognition. The findings pave the way for future investigations into hybrid models, harnessing the strengths of both quantum and classical paradigms to advance predictive capabilities in various domains.

# 10. Appendix

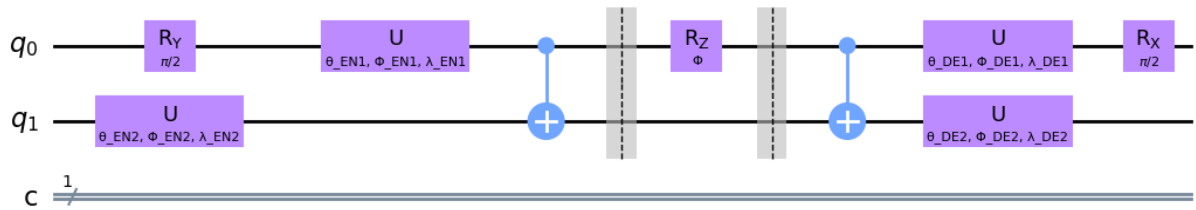

Fig. 2. Parameterized Quantum Circuit of 2 Qubits (Source:Qiskit Lab)

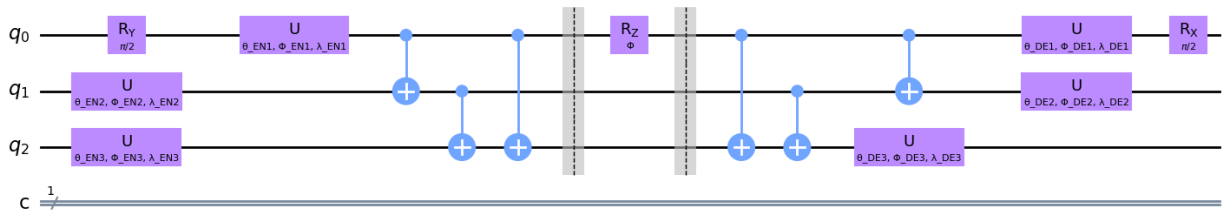

Fig. 2. Parameterized Quantum Circuit of 3 Qubits (Source:Qiskit Lab)

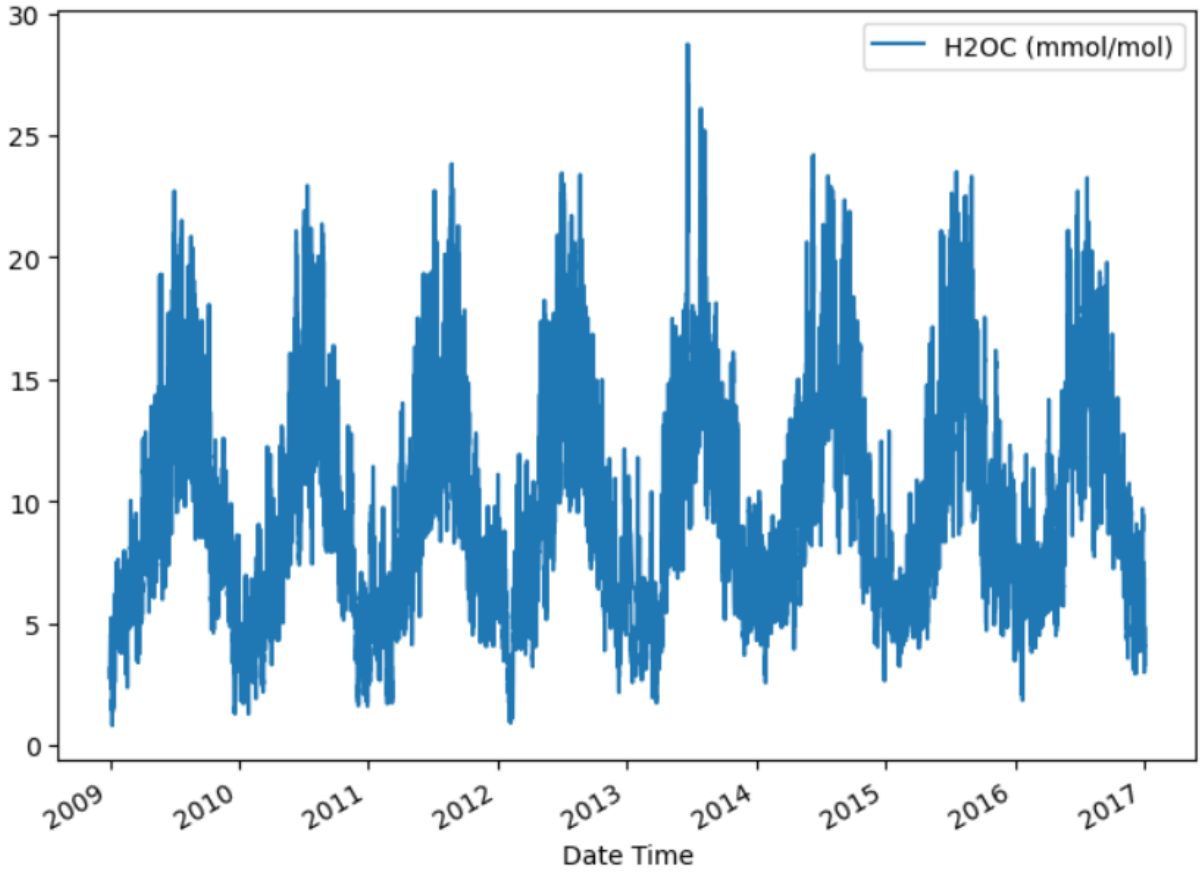

Fig. 3. Time series plot of concentration of Water Vapour $(H2OC)(mmol/mol)$

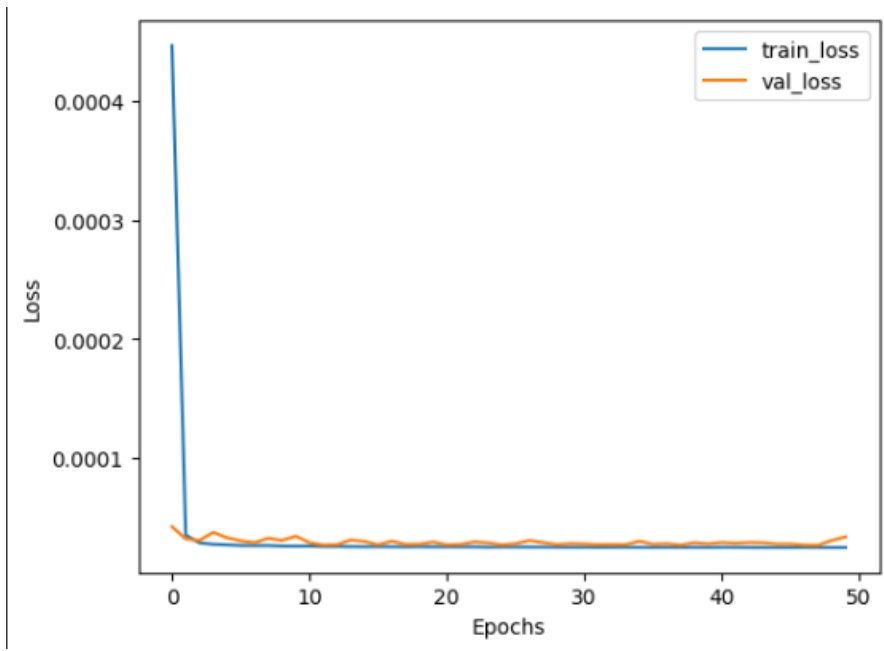

Fig. 4. LSTM loss plot

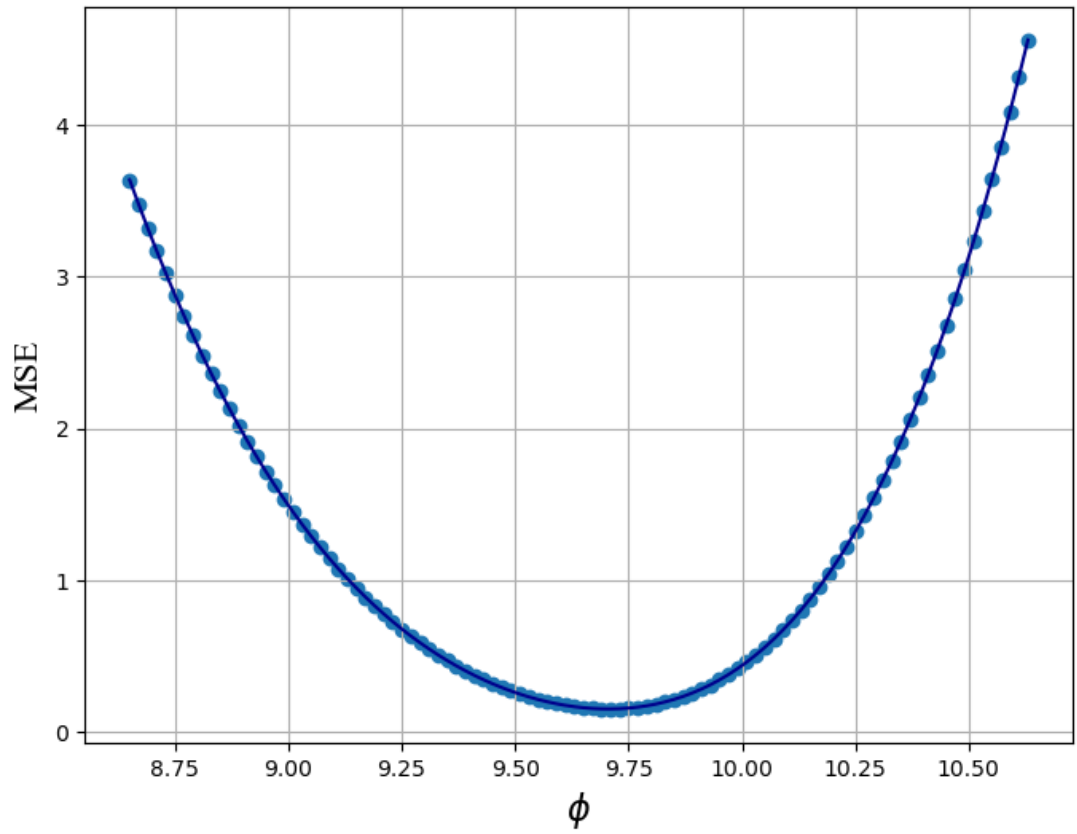

Fig. 5. MSE vs $\phi$ plot for 2 Qubits Circuit, Learning Rate $= 0.01$

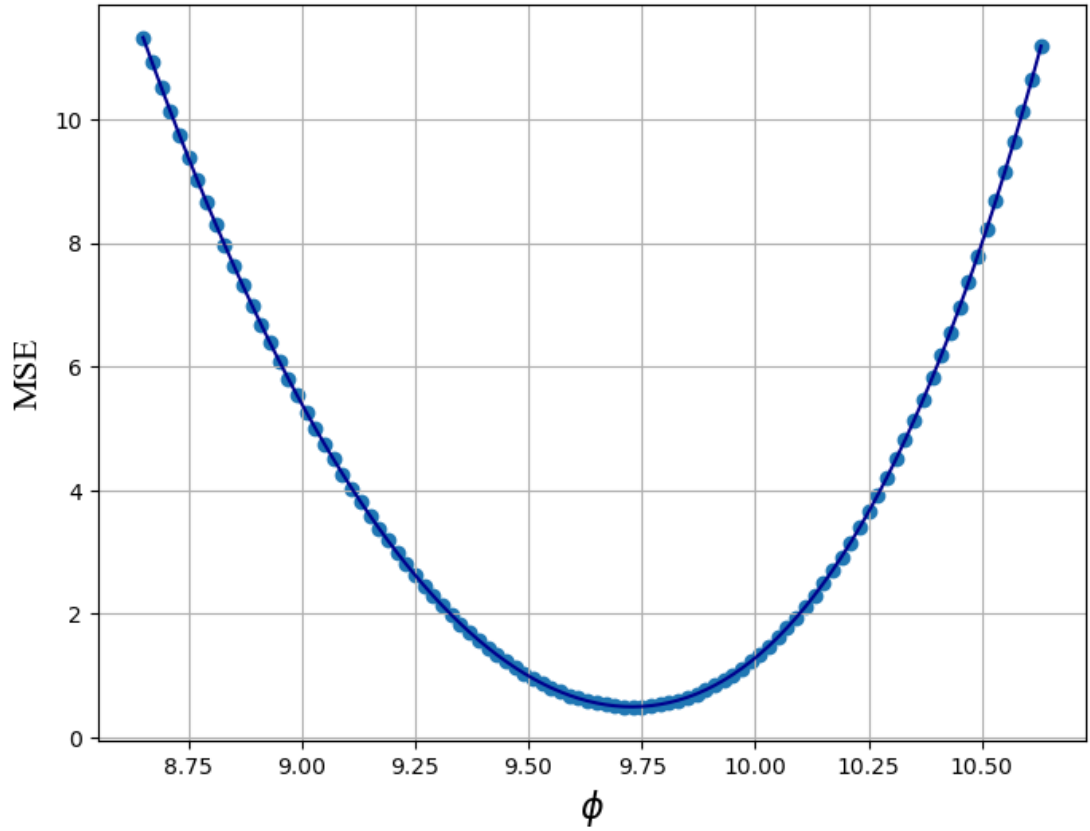

Fig. 6. MSE vs $\phi$ plot for 2 Qubits Circuit, Learning Rate=0.02

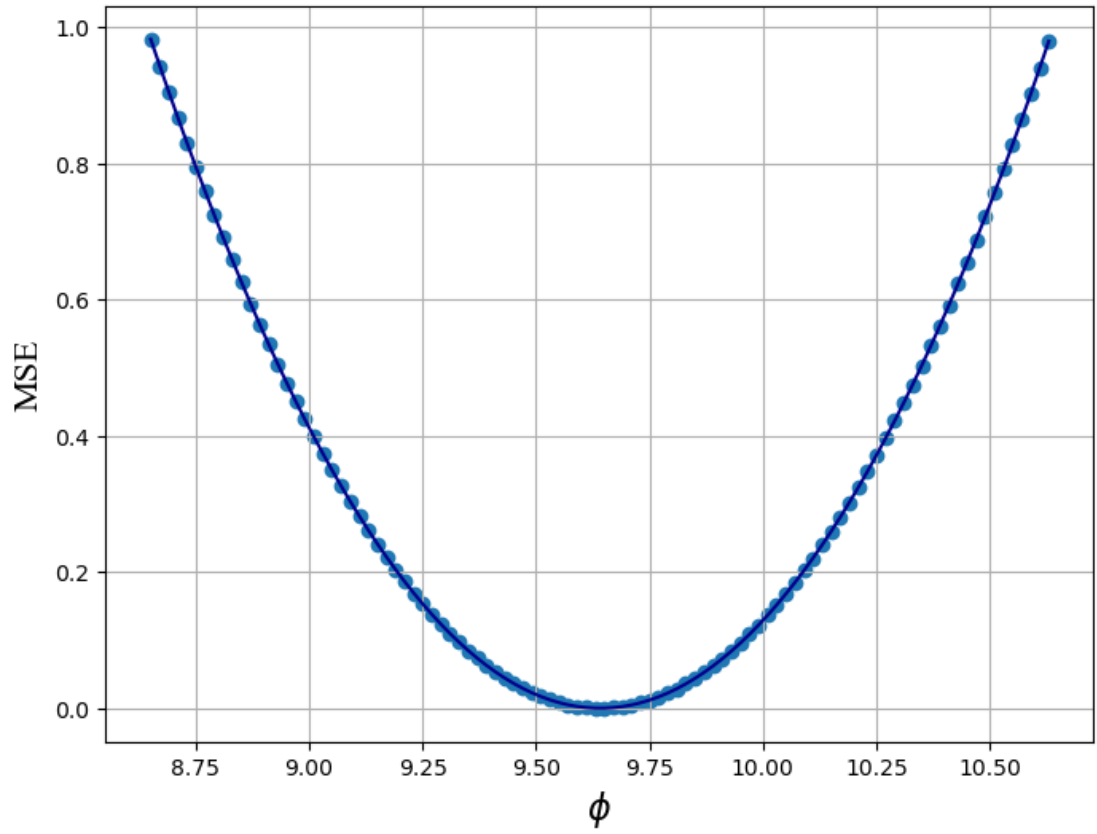

Fig. 7. MSE vs $\phi$ plot for 2 Qubits Circuit, Learning Rate=0.03

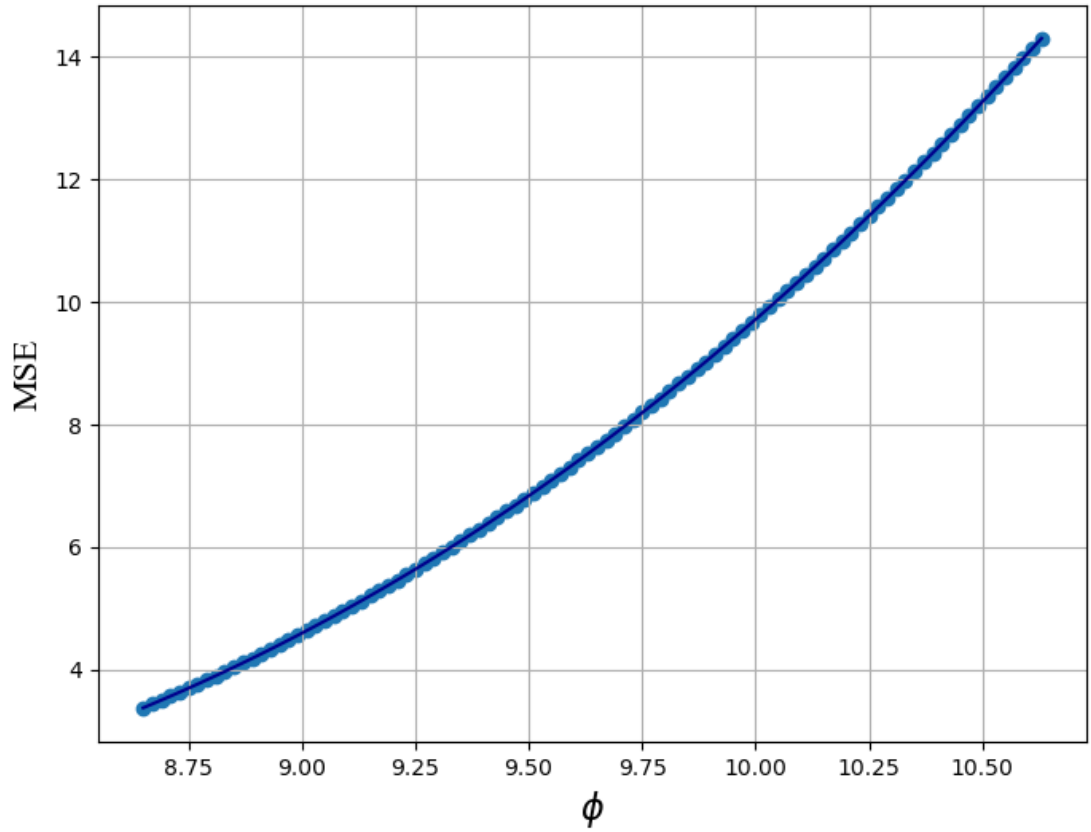

Fig. 8. MSE vs $\phi$ plot for 2 Qubits Circuit,Learning Rate=0.04

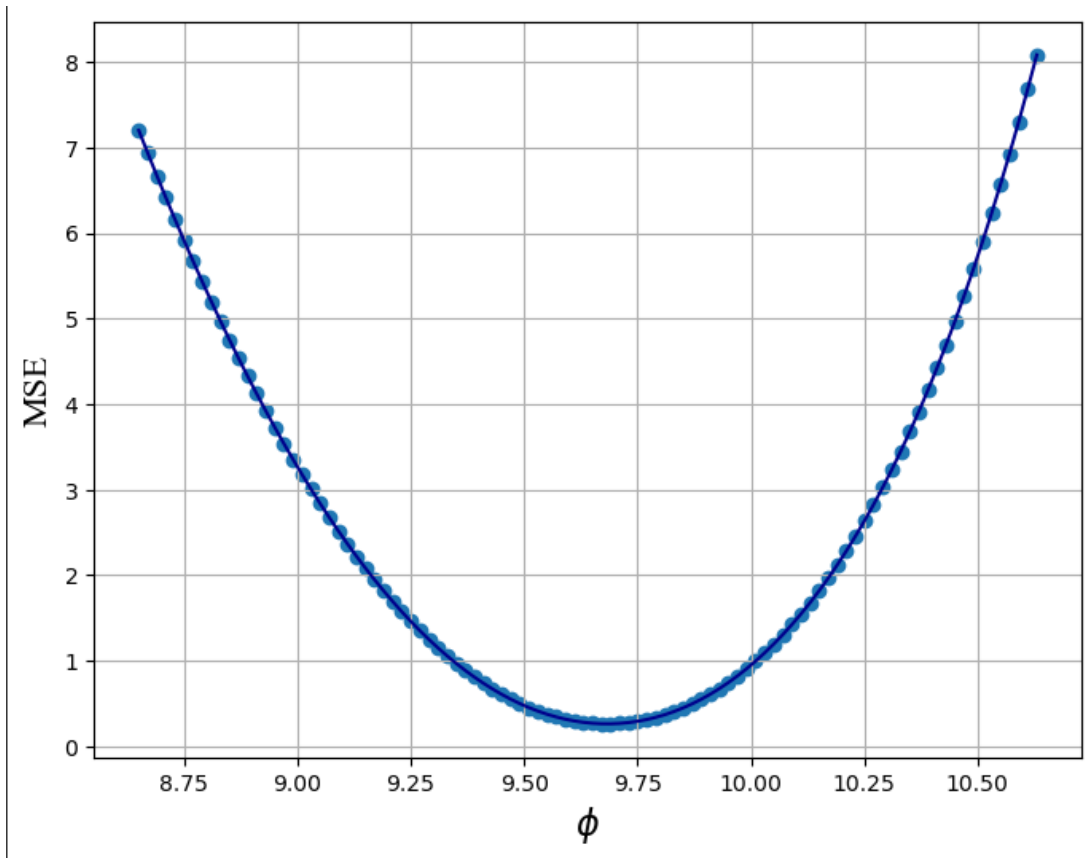

Fig. 9. MSE vs $\phi$ plot for 3 Qubits Circuit,Learning Rate=0.01

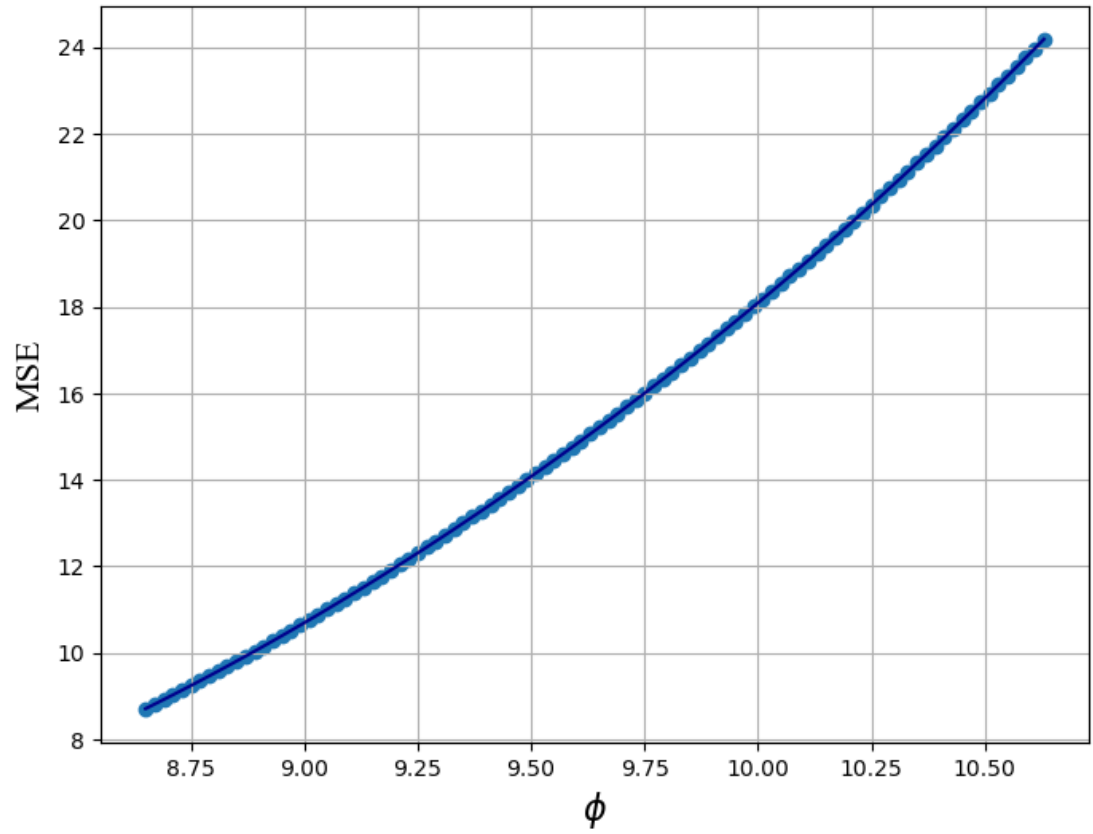

Fig. 10. MSE vs $\phi$ plot for 3 Qubits Circuit, Learning Rate=0.02

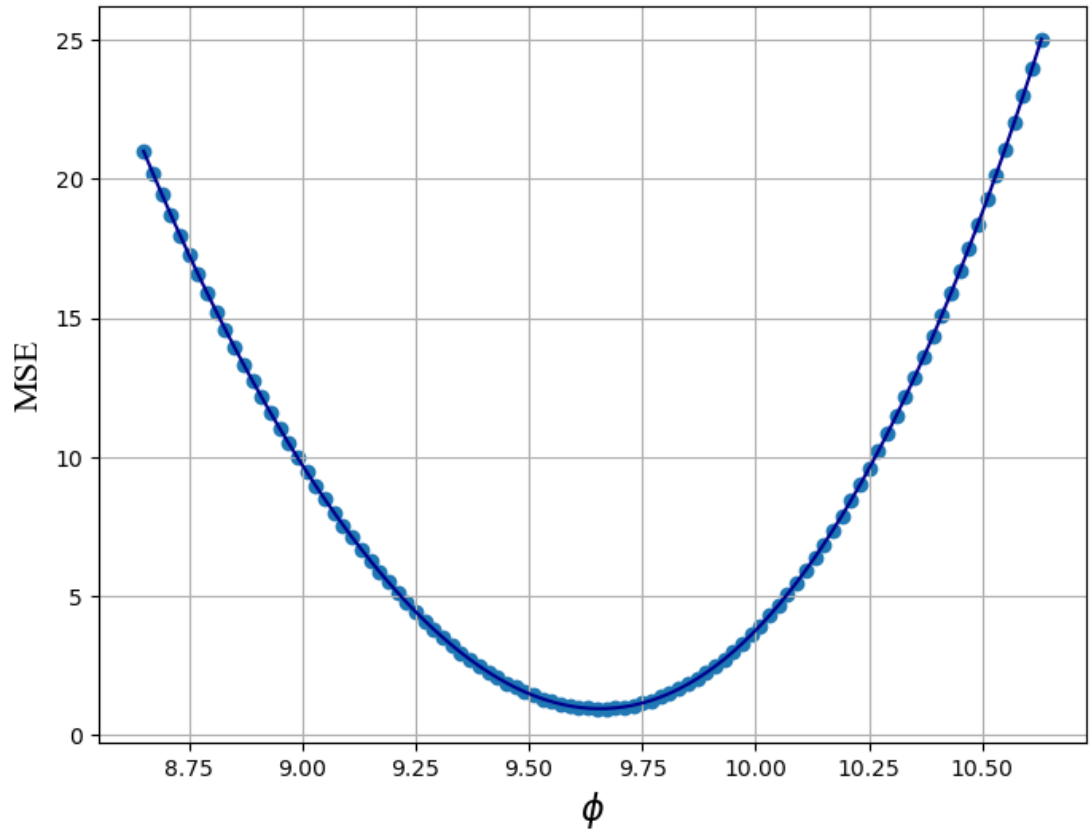

Fig. 11. MSE vs $\phi$ plot for 3 Qubits Circuit, Learning Rate=0.03

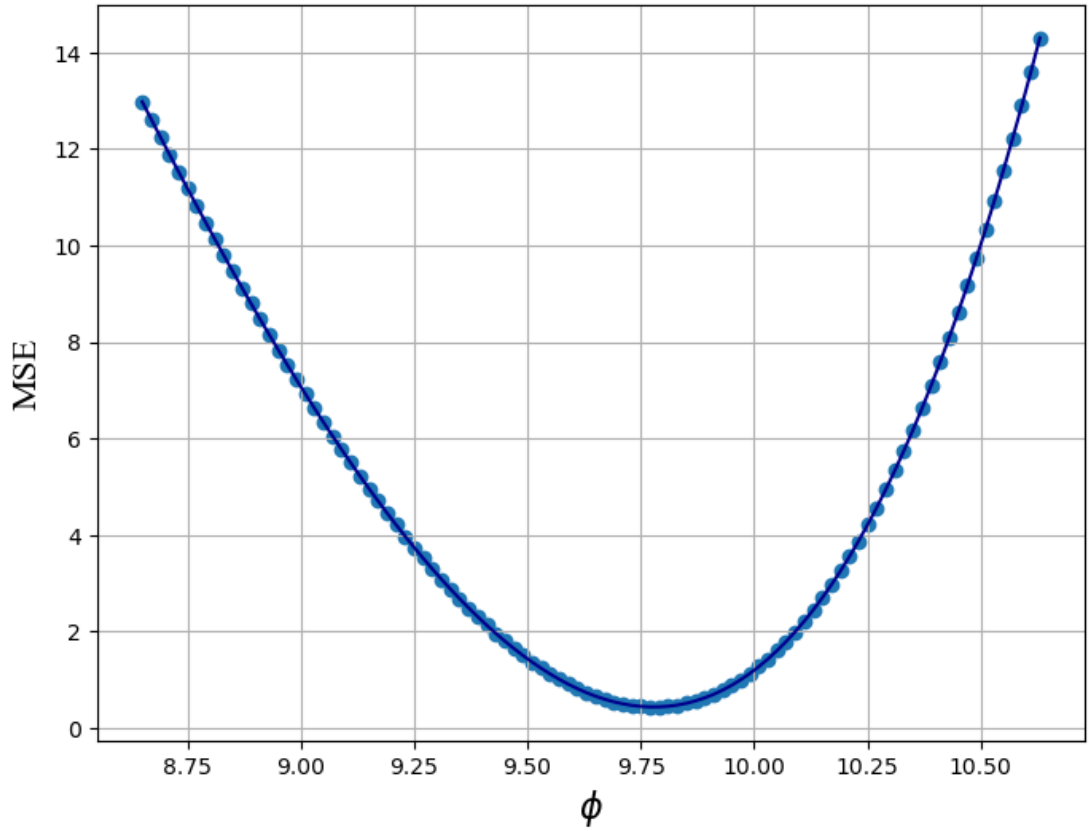

Fig. 12. MSE vs $\phi$ plot for 3 Qubits Circuit, Learning Rate=0.04